\documentclass[12pt]{article} 
\begin{document} 
\begin{center}
          {\large \bf Some remarks about the transition gluon radiation } 

\vspace{0.5cm}                   
{\bf I.M. Dremin}

\vspace{0.5cm}              
          Lebedev Physical Institute, Moscow 119991, Russia

\end{center}

\begin{abstract}
The electrodynamical formulae for the transition radiation are applied to
the case of partons crossing the surface of the deconfined region in 
heavy ion collisions. The chromopermittivity is used in place of the
dielectric permittivity. The corresponding energy-angular distributions of
emitted gluons are discussed. They could be of interest at LHC energies.
\end{abstract}

In 1945 V.L. Ginzburg and I.M. Frank proposed the idea about the transition
radiation \cite{gfra}. Any electric charge passing from some medium to another 
one or
moving with (almost) constant velocity in inhomogeneous medium with variable 
dielectric permittivity induces its polarization and radiation of photons. 
This property has been used, e.g., to build the transition radiation tracker
in ATLAS detector at the LHC. The transition radiation is determined by 
restructuring of the electromagnetic field surrounding the electric charge.

During hadron (nucleus) collisions the partons intersect the surface of 
another partner. This transition should induce abrupt changes of their color 
fields and, consequently, radiation of gluons analogously to electrodynamical
situation. The classical in-medium QCD 
equations coincide in general features with electrodynamical equations 
\cite{inmed} with replacement of the dielectric permittivity by the
chromopermittivity. Since this approach happens to be successful with Cherenkov 
gluons \cite{dklv} and wakes \cite{holz, ryaz, mpl} it is tempting to apply the 
Ginzburg-Frank formula to estimate the distribution of "transition" gluons 
emitted per unit length within some frequency and angular intervals as
\begin{eqnarray}
\frac {dN}{d\omega d\Omega } =
\frac {\alpha _S \sqrt {\epsilon _a} \sin^22\theta}{\pi \omega }\times 
\hspace{8cm}  \nonumber \\
\times \left \vert \frac {(\epsilon _b - \epsilon _a)(1-\epsilon _a+\sqrt 
{\epsilon _b - \epsilon _a \sin ^2\theta})}{(1-\epsilon _a\cos ^2\theta)(1+\sqrt 
{\epsilon _b - \epsilon _a \sin ^2\theta})(\epsilon _b\cos \theta + 
\sqrt {\epsilon _a\epsilon _b-\epsilon _a^2 \sin ^2\theta})}\right \vert ^2.
\label{gfra}
\end{eqnarray}  
$\epsilon _{a, b}$ are the chromopermittivities of the media $a, b$
as felt by relativistic ($v\approx c$) partons traversing the surface when 
passing from $b$ to $a$.  Casimir factors are omitted.  

There is important difference between Cherenkov and transition radiations.
Cherenkov radiation can be observed only if the particle moves in the medium
with $\epsilon >1$. Its intensity is proportional to 
$\Delta \epsilon = \epsilon -1$. Transition radiation is proportional to 
$(\epsilon _b - \epsilon _a)^2$ (see (\ref{gfra})) and appears at any change 
of $\epsilon $. In electrodynamics it is studied mostly in the frequency
interval where the formula
\begin{equation}
{\rm Re }\epsilon _{ED}=1-\frac {\omega _L^2}{\omega^2}         \label{eed}
\end{equation}
is applicable i.e. where $\Delta \epsilon <0$. Here $\omega _L$ is the 
Langmuir (plasma) frequency. Namely this behavior of the dielectric permittivity 
is responsible for specific features of the transition radiation. 

In the case of the nuclear medium it is unknown whether the analogous formula 
may be used in any energy region. For narrow energy intervals of secondary 
particles (as those studied at RHIC) it is reasonable to assume 
$\Delta \epsilon $ to be constant \cite{dklv}. At very high energies the 
dependence 
\begin{equation}
{\rm Re }\epsilon = (1+\frac {\omega _0^2}{\omega ^2})
\Theta (\omega -\omega _{th})   \label{eom}
\end{equation}
may be used. As explained in \cite{drem1, dhe}, it is inspired by the behavior 
of the real parts of the forward scattering amplitudes where they
become positive for all hadrons at $\omega >\omega _{th}$. 

Both these assumptions find some support in RHIC \cite{dklv} and cosmic ray data
\cite{apan, d1, d0}. That is why we consider first these two possibilities. 

The away-side parton at RHIC is created inside the quark-gluon medium, moves
through it and escapes in the vacuum being hadronized. It induces collective
excitation of the quark-gluon medium leading to the double-humped events
(early experimental results see in \cite{fw, adl}). According to fits \
cite{dklv} of 
experimental data $\epsilon _b \approx 6\gg \epsilon _a=1$. Using (\ref{gfra}) 
for estimates of possible transition gluon radiation one 
gets the Bremsstrahlung-like spectrum with both infrared and collinear
divergencies in asymptotics
\begin{equation}
dN\propto \frac {d\omega }{\omega }\frac {d\theta }
{\theta }.     \label{sle}
\end{equation}
Thus, nothing spectacular happens. As usually,
at finite energies the angular divergence is replaced by a maximum at 
$\theta \approx \gamma ^{-1}$. Anyway, this radiation is collimated near the
initial direction of the away-side parton and is hard to detect.
The similar conclusions are obtained \cite{kleo}
for the radiation by a gluon moving in a stochastic medium. It is interesting 
to note that for $\epsilon _b \gg \epsilon _a$ the intensity of 
the transition radiation (\ref{sle}) does not depend on $\epsilon _b$.

For the forward moving partons with extremely high energies 
$\omega >\omega _{th}$ the situation is different. They might feel the slight
excess of the chromopermittivity over 1 according to (\ref{eom}). 
Its specific $\omega $-dependence leads to the spectacular energy
distribution
\begin{equation}
dN\propto \frac {d\omega }{\omega ^5}\frac {d\theta }
{\theta }\Theta (\omega -\omega _{th}).     \label{she}
\end{equation}
The collinear divergence is the same but the energy spectrum has a strong peak 
near $\omega _{th}$. It implies that the resonance-like almost monochromatic 
subjets with masses determined by $\omega _{th}$ will be produced at LHC 
energies. These "resonances" are mostly created by gluons (especially at LHC) 
and, therefore, are neutral. The most favored channel would be to observe 
them as peaks in mass spectra of forward moving $\mu ^+\mu ^-$-pairs near 
$\omega _{th}$. The model cutoff (\ref{she}) at $\omega _{th}$ is replaced
in real data about hadronic amplitudes by some threshold behavior \cite{d0}. 
Unfortunately, there exists no clear prescription how to translate the 
threshold energy in hadronic reactions to gluons with account of hadronization
effects. If observed, peaks locations and their disappearence at 
$\omega = \omega _{th}$ due to the threshold would tell us about that. 

One may hope that the confinement and hadronization do not spoil these 
conclusions. This is supported by observation of effects due to Cherenkov 
gluons and the wake albeit the transition radiation might be stronger influenced by
confinement near the QGP surface than the collective excitations inside 
the volume responsible for Cherenkov gluons and wake radiation.
From one side, it shortens the radiation length but, from another side, 
the abrupt variation of color fields surrounding partons when they come out of 
the quark-gluon plasma and hadronize may enlarge the transition radiation 
because of strong confining forces. At the same time the transition radiation 
may appear in much wider energy intervals than Cherenkov radiation (in 
particular, at $\omega < \omega _{th}$). However, neither experimental nor 
theoretical arguments about the chromopermittivity in these regions exist
nowadays. It would not be surprising if we find its 
effects outside the regions where the double-humped events were observed.
That may somewhat smooth down the structures near $\omega _{th}$.

With all above precautions, I present these remarks to arXiv with hope to 
initiate the discussion of these very preliminary ideas if any signatures 
will be found at LHC energies.

This work was supported by RFBR grants 09-02-00741; 08-02-91000-CERN and
by the RAN-CERN program.

\end{document}